\documentclass[graybox, envcountchap, authoryear, natbib]{svmult}

\usepackage{mathptmx}       
\usepackage{helvet}         
\usepackage{courier}        
\usepackage{type1cm}        
\usepackage{makeidx}         
\usepackage{graphicx}        
\usepackage{multicol}        
\usepackage[bottom]{footmisc}

\def \aap{A\&A}

\def \aj{AJ}

\def \apjl{ApJ}

\def \apj{ApJ}

\def \mnras{MNRAS}

\usepackage{amsmath}
\usepackage{amssymb}
\newcommand{\CII}{\hbox{{\rm C}\kern 0.1em{\sc ii}}}
\newcommand{\CIV}{\hbox{{\rm C}\kern 0.1em{\sc iv}}}
\newcommand{\Fe}{\hbox{{\rm Fe}}}
\newcommand{\FeI}{\hbox{{\rm Fe}\kern 0.1em{\sc i}}}
\newcommand{\FeII}{\hbox{{\rm Fe}\kern 0.1em{\sc ii}}}

\newcommand{\SiII}{\hbox{{\rm Si}\kern 0.1em{\sc ii}}}
\newcommand{\Al}{\hbox{{\rm Al}}}
\newcommand{\AlII}{\hbox{{\rm Al}\kern 0.1em{\sc ii}}}
\newcommand{\AlIII}{\hbox{{\rm Al}\kern 0.1em{\sc iii}}}
\newcommand{\NiII}{\hbox{{\rm Ni}\kern 0.1em{\sc ii}}}

\newcommand{\CrII}{\hbox{{\rm Cr}\kern 0.1em{\sc ii}}}

\newcommand{\MnII}{\textsc{{\rm Mn}\kern 0.1em{\sc ii}}}
\newcommand{\TiII}{\textsc{{\rm Ti}\kern 0.1em{\sc ii}}}
\newcommand{\Zn}{\textsc{{\rm Zn}}}
\newcommand{\ZnII}{\textsc{{\rm Zn}\kern 0.1em{\sc ii}}}

\newcommand{\NaD}{\textsc{[{\rm Na}~D}]}
\newcommand{\NeV}{\textsc{[{\rm Ne}\kern 0.1em{\sc v}}]}
\newcommand{\NII}{\textsc{[{\rm N}\kern 0.1em{\sc ii}}]}
\newcommand{\OIII}{\textsc{[{\rm O}\kern 0.1em{\sc iii}}]}
\newcommand{\OII}{\textsc{[{\rm O}\kern 0.1em{\sc ii}}]}
\newcommand{\OI}{\textsc{{\rm O}\kern 0.1em{\sc i}}}
\newcommand{\MgI}{\textsc{{\rm Mg}\kern 0.1em{\sc i}}}
\newcommand{\MgII}{\textsc{{\rm Mg}\kern 0.1em{\sc ii}}}

\newcommand{\HI}{\textsc{{\rm H}\kern 0.1em{\sc i}}}
\newcommand{\HII}{\textsc{{\rm H}\kern 0.1em{\sc ii}}}
\newcommand{\lya}{\textsc{{\rm Ly}\kern 0.1em$\alpha$}}
\newcommand{\Ly}{\textsc{{\rm Ly}\kern 0.1em$\alpha$}}
\newcommand{\Ha}{\textsc{{\rm H}\kern 0.1em$\alpha$}}
\newcommand{\Hb}{\textsc{{\rm H}\kern 0.1em$\beta$}}
\newcommand{\Hg}{\textsc{{\rm H}\kern 0.1em$\gamma$}}
\newcommand{\SII}{\hbox{[{\rm S}\kern 0.1em{\sc ii}}]}

\newcommand{\mpy}{\hbox{M$_{\odot}$~yr$^{-1}$}}

\newcommand{\msun}{\hbox{M$_{\odot}$}}

\newcommand{\NHI}{\ensuremath{N_\textsc{h\scriptsize{\,i}}}}
\newcommand{\NH}{\hbox{$N_{\rm H}$}}

\newcommand{\kpc}{\hbox{$h^{-1}$~kpc}}

\newcommand{\kms}{km~s$^{-1}$}
\newcommand{\nn}{\nonumber}
\newcommand{\EW}{\hbox{$W_{\rm r}^{\lambda2796}$}}

\newcommand{\Ms}{\hbox{$M_{\star}$}}
\newcommand{\Mg}{\hbox{$M_{\rm gas}$}}

\newcommand{\logZ}{\log Z/Z_{\odot}}
\newcommand{\arcsec}{~{arcsec}}
\newcommand{\degree}{$^{\circ}$} 
\newcommand{\cmsqa}{cm$^{-2}$} 

\newcommand{\sfrJ}{$4.7\pm2.0$~\mpy}

\newcommand{\sfrHE}{$33^{+40}_{-11}$~\mpy}

\begin{document}

\title*{Gas Accretion and Star-Formation Rates with IFUs and Background Quasars} 
 \titlerunning{Gas accretion with IFUs}
\author{Nicolas F. Bouch\'e}
\institute{Nicolas F. Bouch\'e \at IRAP, 9 Av. Colonel Roche, F-31400 Toulouse, \email{nicolas.bouche@irap.omp.eu}
}
%
%
\maketitle


\abstract{Star forming galaxies (SFGs) are forming stars at a regular pace, forming the so-called main sequence (MS).  However, all studies of their gas content show that their gas reservoir ought to be depleted in 0.5-2 Gyr.
 Thus, SFGs  are thought to be fed by the continuous  accretion of intergalactic gas
 in order to sustain their star-formation activity.
However,  direct observational evidence for this accretion phenomenon has been elusive. 
Theoretically, the accreted gas coming from the intergalactic medium is expected to orbit about the halo,
delivering not just fuel for star-formation but also angular momentum  to   the galaxy.
This accreting material is thus expected to  form a gaseous structure that should be co-rotating with the host once at $r<0.3\;R_{\rm vir}$ or $r<10-30$ kpc. Because of the rough alignment between the star-forming disk and this extended gaseous structure, 
the accreting material can be  most easily detected with the combination of background quasars
and integral field units (IFUs).  
In this chapter, accretion studies using this technique are reviewed.
}

\section{Gas accretion in the context of galaxy evolution}
\label{section:boucheintro}

Two major advances in our understanding of the evolution of galaxies have occurred in the past five years or so.
First, star forming galaxies (SFGs) are seen to be forming stars at a regular pace, forming the so-called main-sequence
\citep[MS;][]{BrinchmannJ_04a,NoeskeK_07a,DaddiE_07a,ElbazD_07a,PengY_10a,WhitakerK_12a,WhitakerK_14a,Tomczak2016}
an almost linear relation between star-formation rate (SFR) and stellar mass (\Ms).
Second, star-forming galaxies have  gas consumption time scales (defined as the time scale to consume the gas at the current SFR, i.e. 
\Mg/SFR) that are too short  compared to the age of the Universe 
\citep[at all epochs where gas masses can be measured, currently up to redshifts $z=2$;][]{LeroyA_08a,GenzelR_10a,TacconiL_13a,FreundlichJ_13a,SaintongeA_13a,SaintongeA_16a}. 
Thus, in order to sustain the observed levels of star-formation over many Gyrs,
galaxies must continuously replenish their gas reservoir with fresh gas 
accreted from the vast amounts available in the intergalactic medium.


Two additional indirect arguments point towards the necessity for gas accretion.
One other argument comes from the metallicity distribution of sun-like stars (G-type) in the Milky Way, where the metallicity distribution of G-stars is not consistent with  
the expectation from chemical evolution models (whose initial condition is that of a pure gaseous reservoir, a.k.a. a `closed-box')
  unless some fresh gas infall is invoked  \citep{Lynden-BellD_75a,PagelB_75a}, a discrepancy often referred to as the G-dwarf problem  \citep{vandenberghS_62a,SchmidtM_63a}.
  Another another indirect argument for continuous replenishment of galaxy reservoirs comes from the
  non-evolving (or very weakly evolving) cosmic neutral density $\Omega_{\HI}$ for damped \Ly\ absorbers  \citep{PerouxC_03a,BauermeisterA_10a,NoterdaemeP_12b,ZafarT_13a,CrightonN_15a}
 which contrasts with the rapid evolution of the stellar cosmic density. 
 
In numerical simulations, accretion of intergalactic gas (via the cosmic web) originates from the growth of dark matter halos which pulls the cold baryons along.  In galaxies with luminosities less than $L^*$, this process is expected to be very efficient 
owing to the short cooling times in these halos \citep{WhiteS_91a,BirnboimY_03a,KeresD_05a,FaucherG_11b}.
This phenomenon is often referred to as `cold accretion' and this term describes the halo mass regime ($M_{\rm DM}<M_{\rm crit}$) where the accretion is most efficient \citep{WhiteS_91a,BirnboimY_03a}. 
At high masses and high redshifts, cold accretion can occur in the form of cold-streams \citep{DekelA_09a}, i.e. when the halo is larger
than the typical filament cross-section. 

Once inside the galaxy dark matter halo, the accreted gas is expected to orbit  the galaxy,  
the cold accreted gas should orbit about the halo before falling in to build the central disk, delivering not only fuel for star formation but
 also angular momentum to shape the outer parts of the galaxy  \citep{StewartK_11b,ShenS_13a,DanovichM_15a}. Thus, accreting material should co-rotate with the central disk in the form of a warped, extended cold gaseous structure \citep{PichonC_11a,KimmT_11b,DanovichM_12a,DanovichM_15a,ShenS_13a}, 
 sometimes referred to as a
 ``cold-flow disk'' \citep{StewartK_11a,StewartK_13a,StewartK_16a}. This is further discussed in the chapter by K. Stewart. 

In the local universe, such large gaseous disks are often seen around galaxies in \HI\ 21cm surveys,
where the \HI\ disk extends 2--3 times beyond the stellar radius as in  the M33 low surface brightness disk \citep{PutmanM_09a},  the more massive  M81 \citep{YunM_94a} and M83 galaxies
\citep{HuchtmeierW_81a, BigielF_10a}, and others 
(see chapter by F. J. Lockman). 
The kinematics of this \HI\ gas in the outer parts show that the gas is systematically
rotating in the same direction as the central object at those large radii (2--3 times beyond the stellar radius or 20-30 kpc).
 
As discussed in \citet{StewartK_11a}, these gaseous disks (or ``cold-flow disks'') should produce distinct kinematic signatures in absorption systems.  In particular,
the gas kinematics seen in absorption spectra are expected to be offset by about $100$~\kms\ from the galaxy's systemic velocity.
  It should be noted that these cold-flow disks are the end product of cold accretion and  these two terms should not be confused.
  Moreover, cold-flow disks occur on scales of a few tens of $R_{\rm vir}$ whereas cold accretion (or cold streams) occur on scales of Mpc,
  hence the possibility to  observe with background quasars these two distinct phenomenon depends strongly on the gas column density.
  Theoretical expectations for detecting cold accretion with Lyman limit systems is discussed in \citet{FumagalliM_11a,GoerdtT_12a} and in 
the chapter by C. A. Faucher-Gigu\`ere. 
Observational signatures of cold accretion with Lyman limit systems are discussed in the chapter by N. Lehner. 
This chapter focuses on the distinct absorption kinematics of ``cold-flow disks'' which should occur in the high column density regime 
 and well inside the halos of galaxies.

 Before discussing the most recent observations, it is important to quantify how much accretion is needed  in order to sustain the observed levels of star-formation over many Gyrs
to the levels of tens to hundreds of solar masses per year at $z=1-2$ required by the main-sequence.
This  can be estimated from the following argument.
 As shown in \citet{BoucheN_10a} and many others since \citep{DaveR_12a,KrumholzM_12a,LillyS_13a,FeldmannR_13a,FeldmannR_15a,DekelA_14a,ForbesJ_14a,PengY_14b}, galaxies at $z<4$ can be thought of as  a simplified gas regulator where there is a rough balance between the SFR and the gas accretion rate. 
The equilibrium between gas accretion and SFR under the `regulator'   model  \citep{LillyS_13a} or  `bathtub'  scenario \citep{BoucheN_10a}   
indicate that the accretion rate $\dot M_{\rm in}$ should be comparable to the star-formation rate (SFR).

 In particular, the  equilibrium solution  can be found from the continuity equation between the gas supply and gas consumption terms,  
 and equilibrium SFR can be expressed as
\begin{eqnarray}
\hbox{SFR}&\approx&\frac{\dot M_{\rm in}}{1+R+\eta} \nn\\
&=&\frac{\epsilon_{\rm in}f_{\rm B}}{1+R+\eta}\dot{M}_{\rm h}\label{eq:epsin}
\end{eqnarray}
where $R$ is the gas returned fraction (from massive stars) under the instantaneous recycling approximation, $\eta$ the loading factor from galactic winds, $f_B$ the universal baryonic fraction and $\epsilon_{\rm in}$ the efficiency of accretion (with $\dot M_{\rm in}\equiv \epsilon_{\rm in}\;f_B\;\dot M_{\rm h}$) and $\dot M_{\rm h}$ the dark matter growth rate  \citep{GenelS_08a,McBrideM_09a,DekelA_13a}. 
Equation~\ref{eq:epsin} has several implications.
First, it shows that SFR and the accretion rate $\dot M_{\rm in}$ are comparable to a factor of order unity given that $\eta$ is found to be $\approx1$ \citep{HeckmanT_15a,SchroetterI_15a,SchroetterI_16a}
 and that $R\sim0.5$ with its exact value depending on the initial mass function.
Second, it shows  that the evolution of the accretion rate $\dot M_{\rm in}(z)$ sets the evolution of the SFR \citep{BoucheN_10a}.
Lastly, it is ultimately  the dark matter growth rate that limits the growth rate of galaxies and the evolution of SFR.

Equation~\ref{eq:epsin} is important because it shows that SFR is regulated by cosmological quantities (in other words it is supply-limited)
 and not necessarily the very local conditions (such as gas mass or gas surface density).  The gas mass will adjust itself such that
 there is a close balance between SFR (consumption) and accretion (supply).   However, as discussed in \citet{BoucheN_10a} and \citet{LillyS_13a}, beyond $z>4$ or $z>6$, the dark matter growth rate is so large that the SFR does not have the time to adjust itself towards the (quasi)-equilibrium solution, and thus this solution
does not apply.  

\section{Detecting Gas Accretion}
\label{section:howto}

Contrary to gas outflows where observational evidence abounds,  direct evidence of gas inflows is notoriously 
difficult to obtain. As described in this book, this situation has begun to change.
One can study gas accretion via Lyman Limit Systems (see chapter by N. Lehner),
studies of galaxy spectra (see chapter by K. Rubin),
UV spectroscopy of quasars (see chapter by P. Richter),
or Lyman alpha emission (see chapter by S. Cantalupo). 
In this chapter, I will focus on the recent breakthroughs on detecting gas accretion which came from recent
technological advancements with integral field units (IFUs).

\subsection{Observational and Technological breakthroughs}

To study gas accretion, one can  take advantage of one recent key observational result established only about 5-6 years ago,
namely that the gas distribution as traced by low-ionization lines (such as \MgII, \NaD) is highly anisotropic (see chapter by G. Kacprzak). 
The first evidence which showed that the circumgalactic gas was anisotropic came from a study of \NaD\ absorption in stacked spectra from a  large sample of $\sim1000$ SDSS galaxies \citep{ChenY_10a}. Their results showed
a strong inclination dependence of the blue-shifted absorption lines.
Had the gas been isotropic, this inclination dependence would not have been 
found.

Another important result came soon after with the investigation of the radial and azimuthal dependence of the low ion \MgII\
in a sample of 4000 intermediate redshift  galaxies with 5000 background galaxies  in the COSMOS survey \citep{BordoloiR_11a}. 
They found that --for blue star-forming galaxies-- there is a strong azimuthal dependence of the \MgII\ absorption within 50 kpc of inclined disk-dominated galaxies,  indicating the presence of a strongly bipolar outflows.
This result was soon confirmed with two key studies at $z=0.1-1$ using background quasars which have the advantage that they
can probe the properties of gaseous halos around individual galaxies 
\citep{BoucheN_12a,KacprzakG_12a}.
While the presence of strong bipolar outflows leaving galaxy disks more or less perpendicularly, i.e. in the direction of the minor-axis, these studies with background quasars  revealed that the low-ionization gas traced by \MgII\ do also trace the outskirts of gaseous disks
when the quasar's location appear to be aligned with the minor axis. Indeed, the azimuthal angle $\alpha$\footnote{$\alpha$ is defined as the angle between the galaxy major-axis and the apparent quasar location.} distribution of background quasars with
\MgII\ appears to be bi-modal with a main peak at $\alpha\sim90$~deg (aligned with the minor-axis)  and at $\alpha\sim0$~deg (aligned with major-axis) 
\citep{BoucheN_12a,SchroetterI_15a,SchroetterI_16a}.

\begin{figure}[t]
\centering
\includegraphics[scale=0.2,bb=500 50 1052 1164]{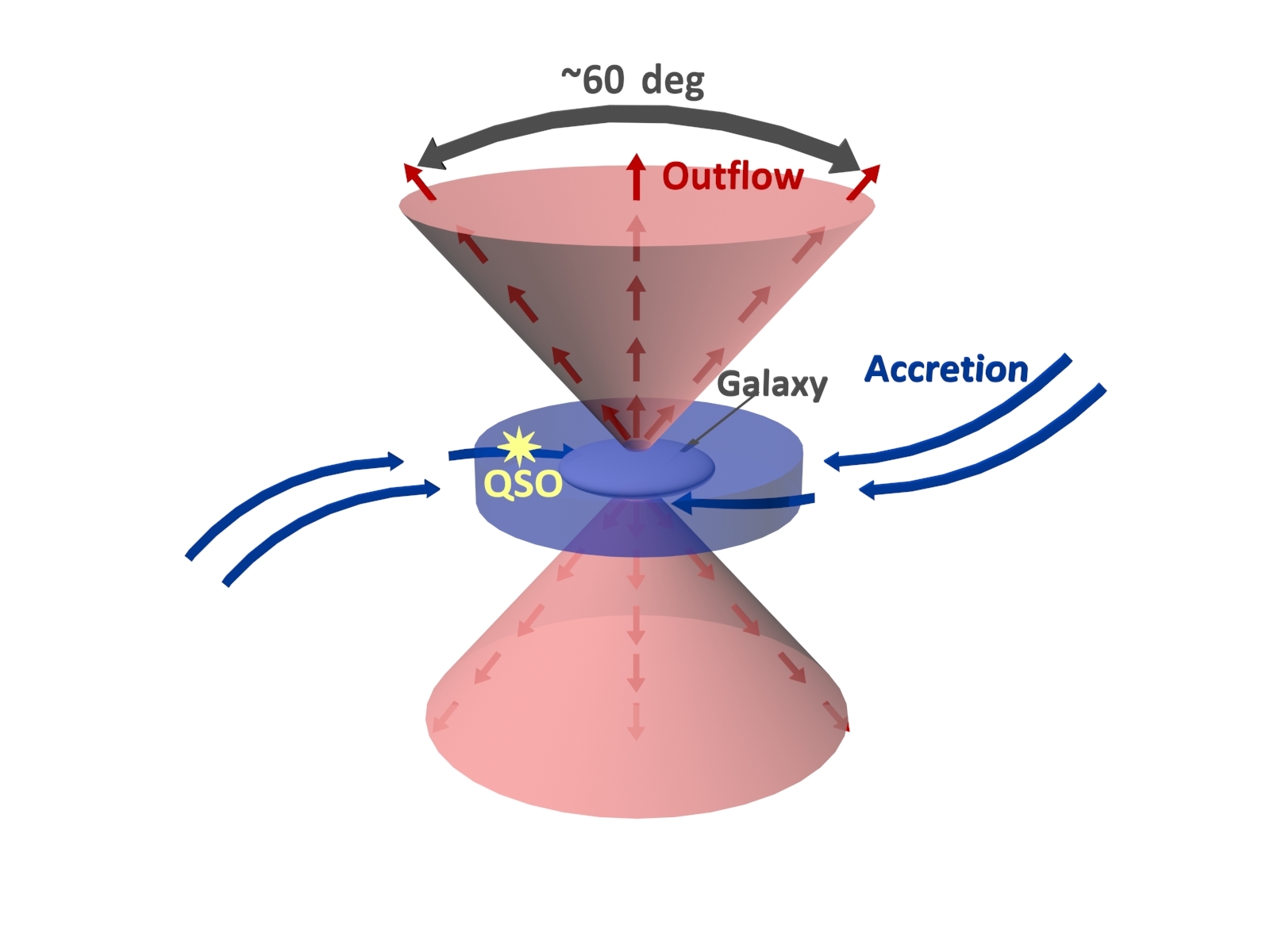}
%
%
\caption{Schematic representation of gas flows around galaxies [reproduced from \citet{BoucheN_13a}]}
\label{fig:gasflows}       
\end{figure}

Thus, a picture is emerging from these observations where outflows occur preferentially along the minor axis of galaxies, 
with also large gaseous
structures (as described in the previous section) that are roughly co-planar with the star-forming galaxy.
Figure~\ref{fig:gasflows} represents schematically the situation with outflows leaving the galaxy along the minor-axis and
gas inflows forming the so-called `cold flow disk', the gaseous structure discussed in the previous section and in the chapter by K. Stewart. 
The cold gas in outflows as traced by \MgII\  extend to at least 80--100 kpc, but outflows could extend to the virial radius.
The cold-flow disk on the other hand is typically 20--30 kpc, i.e. extends to a few tens of the virial radius.

%
   
Quasar absorption lines (such as the low-ionization \MgII$\lambda\lambda$2796,2803 doublet) have had the potential to yield important information on gas flows for decades, but 
the challenge has always been in finding the galaxy associated with a particular absorber.  Indeed, in order to connect the properties of the Circum Galactic Medium (CGM) 
with those of galaxies, such as SFR, mass, etc., it is 
necessary to identify the galaxy associated with each
metal line absorption system. This identification process has been a challenge more decades because it is inherently expensive
and inefficient with traditional imaging and spectrographic instruments. As described in this book, one can
 take deep imaging, but one will need to perform
extensive multi-object spectroscopy of all of the galaxies within a certain impact parameter around the quasar.

\begin{figure}
\centering
\includegraphics[height=10cm,width=10cm]{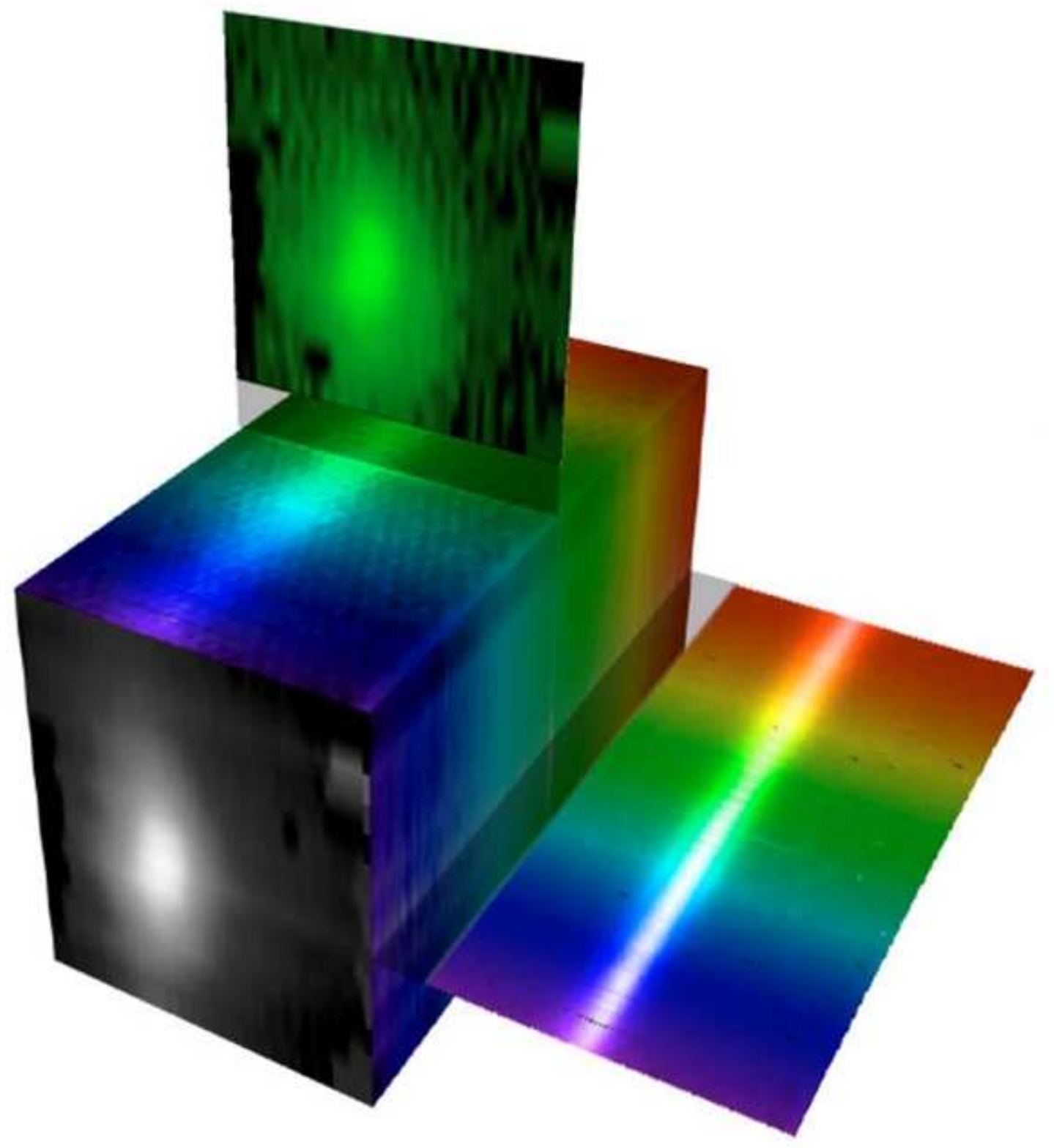}
\caption{Illustration of the advantage in observing quasar fields with IFUs.  A line emitter can easily be
detected at any position in the field of view by selecting the wavelength slice of the hyper-spectral cube corresponding to the expected redshift of the absorber. }\label{fig:ifu}
\end{figure}

But in the past 10 years, there has been a technological breakthrough, with the advancements of IFUs on 10-m class telescopes.
Indeed, with an IFU, one can simply point-and-shoot at a quasar field and identify line emitters at the expected redshift of the absorption system (Figure~\ref{fig:ifu}). However, the IFU must have a sufficiently large field of view (FOV). Two IFUs, both on the VLT, have
played a major role in this area:  the near-infrared IFU SINFONI \citep[with a FOV of 8\arcsec$\times$8\arcsec;][]{BonnetH_04a} and the new optical IFU MUSE \citep[with a FOV of 1'$\times$1';][]{BaconR_10a}.  

This technique was pioneered  by \citet{BoucheN_07a} on a sample of 21 \MgII\ absorbers  selected from the SDSS survey using the SINFONI IFU as illustrated on Figure~\ref{fig:ifu}.  Given that SINFONI has a small FOV of only 8\arcsec$\times$8\arcsec, one has to select the quasar sight-lines with absorbers that have the highest chance to be associated with a galaxy at small impact parameter, i.e. to lie within the FOV.  Thanks to the well-known anti-correlation between the \MgII\ rest-equivalent width and impact parameter \citep{LanzettaK_90a,SteidelC_95b}, the strongest absorbers are the most likely to be associated with a galaxy at a small
(of a few arcsec) impact parameter.  This selection criteria became the basis for the SINFONI \MgII\ Program for Line Emitters 
\citep[SIMPLE; ][]{BoucheN_07a,BoucheN_12a,SchroetterI_15a}.

\subsection{Measuring the Gas Accretion Rate}
\label{section:howmuch}

Suppose one finds a galaxy with a background quasar within 15-30 kpc of a galaxy, but more importantly, a quasar whose
apparent location is aligned with the galaxy's major axis as illustrated on Figure~\ref{fig:gasflows}.  This geometrical configuration is ideal to study the properties of ``cold-flow-disks'' which
are created by the global accretion process (or processes) as discussed earlier. 

The gas accretion rate can be estimated if one approximates the  gaseous structure as a short cylinder of radius $b$ and  height $h_z$ which has mass density $\rho(b)$ at the location of the quasar sight-lines, i.e. at the impact parameter $b$. 
As the cylinder is intercepted by the quasar line-of-sight at the impact parameter $b$,
  the  accretion flux $\dot M_{\rm in}$ through the inner surface of area $A=2\pi b\,h_z$ can be estimated from mass conservation argument, yielding the following equation: 
\begin{eqnarray}
\dot M_{\rm in}(b) &=& 2\pi b\,h_z\,V_{\perp}\,\rho(b) \nn\\
&=& 2\pi b\, V_{\perp}\, \cos(i)\,m_{\rm p}\mu\NH, \label{eq:accrate}
\end{eqnarray}\
 where $V_{\perp}$ is the radial flow velocity (perpendicular to the surface $A$),  $i$ is the inclination of the structure, $\NH$ is the total gas column,
$\mu$ is the mean molecular weight,
$m_{\rm p}$ is the proton mass,
and the second line uses the identity $m_{\rm p}\mu\NH=\int{\rm d}z\rho(b)= \rho(b) h_z/\cos i$.  
This equation can be used to estimate accretion rates of gaseous inflows in the close vicinity of galaxies
as probed by quasar lines-of-sight.

\section{Gas Accretion from IFU surveys }
\label{section:ifus}

In this section, I discuss example results from the recent IFU surveys that have led to the first constraints on the accretion rates of fresh gas around star-forming galaxies.   Before discussing detailed results, I list here several of the key surveys of quasar sight-lines with IFUs. These are
\begin{itemize}
\item  the SINFONI `SIMPLE' survey of 21 sight-lines with strong (rest-equivalent width $>2$ \AA) $z\sim1$ \MgII\ systems \citep{BoucheN_07a,SchroetterI_15a},  which led to a sample of 14 galaxy-quasar pairs, including the J1422 system discussed
in \S~\ref{section:j1422} \citep{BoucheN_16a}; 
\item  the $z=2$ extension of SIMPLE survey  around 18 strong  ($>2$ \AA) $z\sim2$ \MgII\ systems \citep{BoucheN_11a},
 which led to only 4 quasar-galaxy pairs, including the HE2243$-$60 system discussed in \S~\ref{section:he2243} \citep{BoucheN_13a};
\item   the SINFONI survey of quasar sight-lines with strong \HI\ absorbers, with column densities $\log \NHI$ greater than 19-20
\citep{PerouxC_11a,PerouxC_12a,PerouxC_13a};
\item  the MUSE Gas Flow and Wind (MEGAFLOW) survey around \MgII\ systems with rest-equivalent width $>0.8$ \AA,
on 22 quasar sight-lines with multiple ($N=3,4$ or 5) \MgII\ systems \citep{SchroetterI_16a}. 
\end{itemize}

In contrast to the SINFONI surveys where one can only target a single system at a time and have yielded about two dozens galaxy-quasar pairs, the MUSE survey will yield about 80 to 100 galaxy-quasar pairs thanks to the large wavelength range of the MUSE instrument ($4800<\lambda<9300$\AA).

\subsection{Case Study 1: HE2243 at $z=2.32$}
\label{section:he2243}

From a sample of galaxy-quasar pairs, one needs to select suitable pairs for the study of accretion. 
From Figure~\ref{fig:gasflows} one sees that the galaxy should not be seen face-on and that the quasar
apparent location should be aligned with the galaxy major-axis.  
From the z2SIMPLE survey  \citep{BoucheN_11a}, one  star-forming galaxy associated with a $z=2.32$ \MgII\ system is located just 26~kpc from the quasar line-of-sight of HE 2243$-$60~\footnote{This system is also a  damped Lyman absorber \citep[DLA;][]{LopezS_02a} with a column density of $\log \NHI=20.6$.}.
The  background quasar whose apparent location on the sky is fortuitously aligned with the galaxy’s projected major-axis,
makes this galaxy-quasar pair an excellent candidate to test the predictions from cosmological simulations \citep{StewartK_11a,StewartK_16a}.

Not only IFUs are excellent tools to identify the host galaxy, but also, such observations directly yield the galaxy
star-formation rate, the galaxy kinematics and the galaxy's orientation with respect to the quasar sight-line.
In this case, the galaxy has a star-formation rate (SFR) of \sfrHE\ (where $\msun$ is the mass of the Sun), typical for that redshift.
The kinematics    of the $z=2.3285$ star-forming galaxy    obtained with adaptive optics (AO) assisted SINFONI 
(i.e. obtained  with $\sim1$~kpc or $0.25$ arcsec resolution)  reveal that this  galaxy has  physical properties 
 typical for  rotationally-supported disks seen at that redshift \citep{ForsterSchreiberN_06a,ForsterSchreiberN_09a},
 with a maximum rotation velocity of $V_{\rm max}=150\pm15$~\kms\ determined from 3D fitting.
 And the apparent quasar location on the sky is only 10\degree\ to 20\degree\ from the galaxy's projected major-axis.

The kinematics of the absorbing gas along the line-of-sight give clues about the nature of the gas, which were observed with
the VLT spectrograph UVES.   A direct comparison between the absorbing gas kinematics with respect to the galaxy emission kinematics,
reveal that  the gas seen in absorption 26~kpc from the galaxy --- corresponding to 7 times the half-light radius $R_{1/2}=3.6$~kpc or one third of the virial size of the halo $R_{\rm vir}$ ---  is moving in the same direction as the galaxy rotation; i.e. the gas appears to be co-rotating.  The comparison is shown in   Figure~\ref{fig:kine} where panel (a)  shows the galaxy emission kinematics as recovered from the IFU data, and panel (b) shows absorbing gas kinematics measured from the absorption lines in the UVES spectra.
One can see that all of the  low-ionization ions \ZnII, \CrII, \FeII\ and \SiII\ 
tracing the cold gas   show redshifted velocities and 
the observed velocity field of this   galaxy also shows redshifted velocities in the direction of the quasar location.
Hence, the absorbing gas appears to be ``co-rotating'' with the host, meaning that it rotates in the same direction.

However, while the sign of the kinematic shifts agrees, the amplitude of the kinematic differences does not agree entirely.
Only a fraction of the velocities seen in absorption  are in agreement with those seen in emission from the galaxy (extrapolated at the location of the quasar).    This would indicate that a simple rotating disk
with circular orbits and a normal (flat) rotation curve can account only for some fraction of the absorbing gas kinematics.
The other  components  have a line-of-sight velocity less than the rotation speed, indicating that this gas is not rotationally supported. 
 In other words, the second group  of components  is offset  by about 100\kms\ from the rotation pattern;
 hence   cannot be gravitationally supported  and therefore should be flowing in. 
 
 In order to estimate the accretion rate from these observations, one needs estimates of the gas column density 
 and of the inflow speed.  A detailed analysis of the metal column densities in each of the components \citep{BoucheN_13a} indicate that each of these two groups contain about half of the total   \HI\ column  $\log\NHI\simeq 20.6$.
Using a simple geometrical toy model of accretion ---where  ``particles''  representing gas clouds or absorption components are distributed in an extended disk configuration as in
 Figure~\ref{fig:gasflows}---  with cloud kinematics reflecting an inflowing (radial) gas flow, 
 one can generate  simulated absorption profiles which can then be compared to the observed absorption profile.
Because the relative galaxy orientation is well determined
by the VLT/SINFONI data, the only free parameter is the inflow speed which is found to be $\sim200$ \kms.
Figure~\ref{fig:kine}(c) shows the modeled kinematics with the disk component shown in red, and the inflow component shown in black.

Using the constraints on $V_{\rm in}$ and on $\NHI$ from the line-of-sight kinematic shift between the SINFONI and UVES datasets,
the estimated accretion rates estimated from Equation~\ref{eq:accrate} is 30 to 60~\mpy\ 
given the uncertainties in the column density and inflow speed.
The range of accretion rates, 30 to 60~\mpy, is found to be similar to the galaxy's SFR of $\sim$\sfrHE,
 in agreement with  the simplest arguments for galaxy growth given in \S~\ref{section:boucheintro}.
Furthermore, for this galaxy's halo mass,  $M_{\rm h}\sim4\times10^{11}\,\msun$
(assuming $V_{\rm max}\sim V_{\rm circ}$), the maximum baryonic accretion rate
is $f_B\; \dot M_{\rm h}\approx40$ \mpy\ \citep{GenelS_08a,DekelA_13a}, which implies an accretion efficiency  $\epsilon_{\rm in}$ 
in Equation~\ref{eq:epsin} of $\sim100$\%\  regardless of the value for the wind loading factor  $\eta$.


\subsection{Case Study 2: J1422 at $z=0.91$}
 \label{section:j1422}
 
 From the  SIMPLE survey   \citep{BoucheN_07a} consisting of 14 galaxies found around strong $z\sim0.8$--1.0 \MgII\ absorbers  selected from the Sloan Digital Sky Survey (SDSS) database with rest-frame equivalent widths
 $\EW>$2~\AA, 
 two have the galaxy's major-axis roughly aligned with the quasar apparent location 
 among the galaxies that are not seen face-on (with an inclination $>30$\degree) as shown in \citep{SchroetterI_15a}.
 One of these two is the $z=0.9096$  quasar--galaxy pair towards the quasar SDSS J142253.31$-$000149 (hereafter SDSS J1422$-$00),
 where the galaxy's major-axis is about 15\degree\ from quasar-galaxy orientation.
  This is --- as already mentioned --- the most favorable  configuration to look for the kinematic signatures of gas inflows (Figure~\ref{fig:gasflows}).
 
\begin{figure}
\centering
\includegraphics[width=12cm]{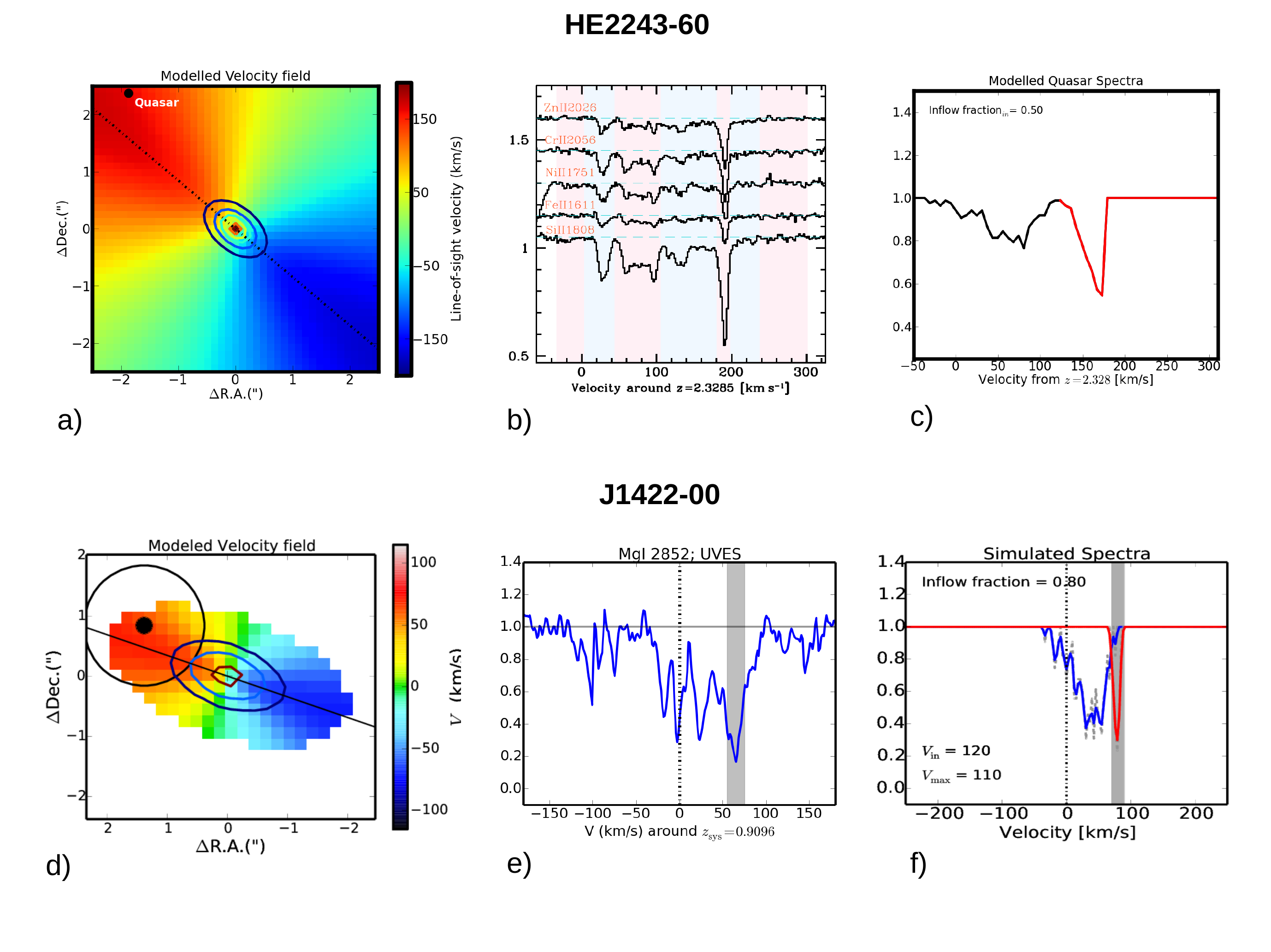}
\caption{Panels a, b and c show, for the HE2243$-$60 galaxy-quasar pair, the galaxy kinematics, the gas kinematics at the quasar location, and the modeled kinematics, respectively. The disk kinematics (with circular orbits) is shown in red while the gas inflow model is shown in black.  Panels  (d) (e) (f) are for the J1422$-$00 pair. }\label{fig:kine}
\end{figure}

 The J1422$-$00 quasar field was also observed with the MUSE IFU \citep{BoucheN_16a}, yielding \OII\ and \Hb\
 fluxes in addition to the \Ha, and \NII\ fluxes from SINFONI.  Incidentally, the MUSE observations  show that there are
   no companion down to 0.22 \mpy\ (5 $\sigma$) within 240 \kpc\ (30\arcsec).
 Combining \Ha, \OII\ fluxes from SINFONI and MUSE observations, the galaxy has a SFR of \sfrJ\ \citep{BoucheN_16a},
 an a metallicity of $12+\log {\rm O/H}=8.7\pm0.2$. Figure \ref{fig:kine} shows the  galaxy kinematics which yield a maximum velocity $V_{\rm max}=110\pm10$ \kms, corresponding to a halo mass of 
$M_{\rm h}\approx2\times10^{11}\,\msun$, corresponding to a 0.1 $L_{\star}$ galaxy. 
 For such a halo mass, the average baryonic accretion rate is
 $f_B\;\dot{M}_{\rm h}\approx 4$ \mpy.

 The \HI\ gas column density associated with this 3.2\AA\ \MgII\ absorber is log\,$N$(\HI)$\simeq20.4$, measured from a
{\it Hubble Space Telescope}/COS G230L spectra.
 Using a high-resolution  spectrum (UVES) of the background quasar, 
 \citet{BoucheN_16a} found that
 the gas kinematics traced by low-ionization lines (\Mg, \Zn, \Fe,   \Al) show distinct signatures consistent
  with those expected for a ``cold-flow disk''.
 Indeed, the absorption profile consists of a component that follows the expected (projected) rotation speed, at $+60$ \kms,
 with several sub-components that are offset kinematically  as shown in Figure~\ref{fig:kine}(b)
 Using a similar geometrical model as in \S~\ref{section:he2243}, the absorption kinematics data are
  consistent with a radial inflow speed of $V_{\rm in}\approx$100~\kms, when comparing the data (panel e) and the model (panel f)
in  Figure~\ref{fig:kine}.
With such an inflow velocity, one can estimate the mass flux rate $\dot M_{\rm in}$ to be at least two to three times larger than the SFR
using Equation~\ref{eq:accrate}.
 
 Interestingly, this galaxy also shows evidence for a galactic  wind  from the blueshifted (by $v_{\rm out}=-80\pm15$  \kms) low-ionization absorption lines (\MgII\ and \FeII) in the galaxy  spectrum. The  estimated mass outflow rate is about 0.5--5 \mpy\ (i.e. a loading factor $\eta\leq1$), and currently, this is the only galaxy with simultaneous constraints on the gas outflow and inflow rates.
 With a constraint on the outflow rate and the loading factor, one can  better  estimate the efficiency of baryonic accretion $\epsilon_{\rm in}$ using   Eq.~\ref{eq:epsin}. Together these numbers imply that the factor $(1+R+\eta)$ in  Eq.~\ref{eq:epsin} is  $(1+R+\eta)\times$SFR is 5 to 8~\mpy.
 As mentioned, for this galaxy halo mass, the {\it average} baryonic accretion rate is estimated to be
 $f_B\;\dot{M}_{\rm h}\approx 4$ \mpy\ using the theoretical expectation for the DM accretion rate
from \citet{GenelS_08a} or \citet{BirnboimY_07a}. 
Putting it together, the accretion efficiency is high at $ \epsilon_{\rm in}\approx1.0$ (regardless of the loading factor $\eta$)
in agreement with the theoretical expectation of \citet{DekelA_06a}.
It is important to keep in mind that our measurements on inflow and outflow rates are instantaneous quantities, while this baryonic estimate is time averaged, or averaged for populations of galaxies.

 \subsubsection{Implications for angular momentum}

A critical question for these gaseous disks around star-forming galaxies is how much angular momentum they carry as discussed in \citet{StewartK_13a} and \citet{DanovichM_15a}.  Here, we recall the common relation between the specific angular momentum $j\equiv J/M$ and  the spin parameter taken from \citet{BullockJ_01a},  which is
\begin{equation}
\lambda\equiv \frac{j}{\sqrt{2}\,R_{\rm vir}\,V_{\rm vir}}, \label{eq:bullock}
\end{equation}
where $R_{\rm vir}$ and $V_{\rm vir}$ are the halo virial radius and virial circular velocity.

The host galaxy is estimated by \citet{BoucheN_16a} to have  $\lambda_{\rm gal}$ to be 0.04--0.05, using the
  the relation between disk scale length $R_{\rm d}$ and the disk spin $\lambda$ parameter from \citet{MoH_98a}.
But the angular momentum of the `cold-flow disk' is harder to estimate without a direct size constraint, but one can place useful limits
using the minimum radius given by the impact parameter.  
Doing so, the  spin parameter of the cold-flow disk $\lambda_{\rm cfd}$ is estimated to be
  $\lambda_{\rm cfd}>0.06$, i.e. 50\%\ larger than that of the galaxy.
This limit on the cold-flow disk angular momentum is consistent with the theoretical expectation of \citet{DanovichM_15a},
where the baryons within $0.3\;R_{\rm vir}$ have 2--3 times the galaxy angular momentum.

\subsubsection{Metallicity aspects}

When discussing gas inflows around galaxies, one naturally wonders whether the accreting material has a metallicity of `pristine' IGM gas (i.e. $\logZ=-2$).   Naturally, estimates of the gas metallicity requires a \HI\ column density measurement in addition to
gas phase column densities for the metallic ions.   
For both of the cases discussed in this section, the \HI\ column density could be determined directly from either a UVES optical spectrum or a COS UV-spectrum. 

In both cases, the metallicity of the absorbing gas is estimated to be  much higher than  the IGM metallicity of $\logZ=-2$ for fresh
infall, implying a significant amount of recycling. Indeed, the absorbing gas metallicity is estimated to be $-0.72\pm0.04$ and  $-0.4\pm0.4$ 
for HE2243$-$60 and J1422$-$00, respectively.  Though, in both cases, the host galaxy ISM metallicity --- measured from its nebular lines ---
is somewhat higher than the absorbing gas, at 1/2 $Z_{\odot}$ and $Z_{\odot}$, respectively. 
Hence, one can conclude from these studies that the gas accretion is already enriched from significant recycling, but is lower than the 
metallicity of the ISM by about a factor of two.

\subsection{Case Study 3: H I selection}

Because a selection using metal lines inevitably leads to a bias in the selection of absorption line systems, 
C. Péroux and her collaborators have led several studies using \HI-selected absorbers, i.e. Damped \Ly\ absorbers (DLAs) 
with \HI\ column densities greater than $2\times10^{20}$ \cmsqa, and sub-DLAs with  \HI\ column densities between $10^{19}$
and $10^{20}$ \cmsqa. The advantage of the \HI-selection is that it bypass the biases present in metal line selections.
The disadvantage of this technique is that the success rate in detecting the host galaxy is much lower.

Using the IFU SINFONI,  out of a large survey of 30  \HI\ systems, only 6 were  detected \citep{PerouxC_11a,PerouxC_12a,PerouxC_13a,PerouxC_14a,PerouxC_17a}.  
Out of these 6 detections they find evidence for the presence of outflows in two systems, while three are consistent
with gas accretion and the remaining case is poorly constrained.

Such sample will prove important in order to compare the gas metallicity seen along the quasar line-of-sight with the metallicity of the host as described in \citet{PerouxC_16a}. At the moment, the situation is unclear and large samples are required.

\section{Future Perspectives}
The case studies described in Section 3 are interesting, but the 
natural question is how common are 
the  features shown in Figure~\ref{fig:kine}.
While SINFONI has played a major role in this area, the wide-field IFU 
MUSE, one of the second generation instrument for the VLTs  which started operation in the Fall of 2014, is changing our view of the CGM and will allow for a major leap forward in our knowledge of gas inflows and outflows. Indeed, thanks to its large 1'x1' field of view sampled at 0.2\arcsec/spaxel, large wavelength range from  480nm to  930 nm sampled at 1.25\AA, and to its unprecedented sensitivity, MUSE \citep{BaconR_10a} will allow us to study gas flows around galaxies with   samples 10 times larger (i.e. $\approx 100$)  than current surveys.
In particular, we started the MusE Gas FLO and Wind (MEGAFLOW) survey designed to yield $\approx$100 galaxy-quasar pairs in about 22 quasar sight-lines selected from the SDSS database (DR12) which have multiple (4 or more) \MgII$\lambda\lambda$2796,2803 absorbers at redshifts $0.3 <z<1.4$ suitable for \OII$\lambda\lambda$3726,3729 emitters.
Results on outflows from the first two fields appeared in \citet{SchroetterI_16a} and the first results on accretion from this survey are to appear in J. Zabl (2017, in prep.).

\end{document}